\documentstyle[11pt,a4]{article}
\input{epsf.tex}
\begin{document}
\begin{center}
{\bf New technique for a simultaneous estimation of the level
density and radiative strength functions of dipole transitions at
$E_{ex}\leq B_n -0.5$ MeV}\\
{ V.A. Khitrov, A.M. Sukhovoj}\\
{Frank Laboratory of Neutron Physics, Joint Institute for
Nuclear Research, 141980 Dubna, Russia}
\end{center}

\begin{abstract}
The new, model-independent method to estimate simultaneously the level
densities excited in the $(n,\gamma )$ reaction and the radiative
strength functions of dipole transitions is developed. The method can be
applied for any nucleus and reaction followed by cascade  $\gamma$-emission.
It is just necessary to measure the intensities of two-step
$\gamma$-cascades depopulating one or several high-excited states and
determine the quanta ordering in the main portion of the observed
cascades. The method provides a sufficiently narrow interval of most
probable densities of levels with given $J^{\pi}$ and radiative strength
functions of dipole transitions populating them.
\end{abstract}

\section{Introduction}
The observed parameters of the cascade $\gamma$-decay of the compound
nucleus can be reproduced in the calculation if one determines (in the
frameworks of some model) at least
\begin{enumerate}
\item the mean density $\rho$ of the excited states with given spin and
parity $J^{\pi}$, and
\item the mean width $\Gamma_{\lambda i}$  of $\gamma$-transitions
between the arbitrary states $\lambda$ and $i$.
\end{enumerate}
The objects of primary interest are the total radiative width
$\Gamma_{\gamma}$ of the compound nucleus (neutron resonance) and the
spectrum of $\gamma$-emission. It may be, for example, the intensity
$I_{\gamma\gamma}$ of the cascades of two successive
$\gamma$-transitions between the compound state and given low-lying
level via a great number of intermediate levels. The experimental data
on $I_{\gamma\gamma}$ (as a function of the energy of their intermediate
level) are obtained for over 30 nuclei from the mass region
$114\leq A\leq 200$ (see, e.g., \cite{Bon99}) with a precision of
approximately 10\%. The experimental values of $\Gamma_{\gamma}$ are
known within the same accuracy. Unfortunately, such accuracy cannot be
achieved in the calculation of these parameters for an arbitrary nucleus
because there are no models that would predict $\rho$ and
$\Gamma_{\lambda i}$ with the mentioned above precision.

This is seen from the data of Table 1 which represent the mean ratio
between the experimental cascade intensities $I_{\gamma\gamma}^{exp}$
and those calculated $I_{\gamma\gamma}^{cal}$ using the known enough
models of level density \cite{Dilg73,Ign75} and radiative strength
functions \cite{Axel62,Kadm83}.
\begin{center}{Table~1: The ratio $R= I^{exp}_{\gamma\gamma}/I^{cal}_
{\gamma\gamma}$ averaged over 30 nuclei.}\end{center}
\begin{center}
\begin{tabular}{|c|c|c|c|c|}\hline
Models: & \cite{Dilg73,Axel62}& \cite{Dilg73,Kadm83}&
\cite{Ign75,Axel62} & \cite{Ign75,Kadm83} \\ \hline
 R      & 2.2(2)& 2.7(2) & 1.5(1) & 1.7(1) \\\hline
\end{tabular}
\end{center}

Precise $\gamma$-decay parameters are, however, necessary for the
calculation of the interaction cross-sections of neutrons with unstudied
target nuclei and the understanding of the behavior of nuclear matter
with increasing excitation energy. An analysis of the existing methods
for the determination of the level density \cite{Svir88,Ign78} and
radiative strength functions ($k$) \cite{Bart73}
\begin{equation}
k=\Gamma_{\lambda i}/(E_{\gamma}^3\times A^{2/3}\times D_{\lambda})
\label{eq:one}
\end{equation}
in deformed nuclei, for example, shows that it is not possible to obtain
sufficiently precise experimental level densities for certain intervals
of their energies and quantum numbers as well as the widths of the
corresponding transitions. Analysis of contributions of different
sources of systematical errors in determination of the level density
from the evaporation spectra was performed by H.Vonach \cite{Vona83}
mainly for light spherical nuclei. The total uncertainty evaluated by
him amounts to about 20-30\%. It should be noted, however, that an
accuracy in calculation of cascade intensities within the models
\cite{Dilg73,Ign75,Axel62,Kadm83}in the lightest nuclei (from the range
of the minimum of the neutron strength function) studied by us is also
considerably better than that for deformed nuclei: the ratio $R$ for
$^{114}Cd$ and $^{124,125}Te$ varies in limits from 0.7 ($^{124}Te$) up
to 1.4 ($^{114}Cd$).

Therefore, without developing new methods for the determination of
nuclear parameters under discussion one cannot expect any progress in
the modification of the existing theoretical models, first of all, for
deformed nuclei. (In eq. (\ref{eq:one}) $\Gamma_{\lambda i}$ is the
partial radiative width of $\gamma$-transition with the energy $
E_{\gamma}$, $D_{\lambda}$ is the average level spacing of the decaying
state and quantity $A$ is the nucleon number).

A new and sufficiently perspective way to obtain such information for
the entire energy interval below $B_n$ seems to be the investigation
\cite{Bon191,Bon291} of the two-step $\gamma$-cascades between the
compound state $\lambda$ and the given low-lying level $f$ through all
possible intermediate states $i$. The algorithms
\cite{Pop84,Bon391,Bon95} developed for the analysis of $\gamma -\gamma$
coincidences registered by ordinary $Ge$ detectors allow one to
determine the intensity distribution of the cascades as a function of
the energy of the cascade intermediate levels over the whole energy
region up to $E_{ex}\simeq B_n$ with an acceptable systematic error
(which decreases as the efficiency of the $\gamma$-spectrometer
increases).

The intensity $i_{\gamma\gamma}$ of an individual cascade is
\begin{equation}
i_{\gamma\gamma}=\frac{\Gamma_{\lambda i}}{\Gamma_{\lambda}}\times
\frac{\Gamma_{if}}{\Gamma_i},
\label{eq:two}
\end{equation}
where $\Gamma_{\lambda i}$ and $\Gamma_{if}$ are the partial widths of
the transitions connecting the levels $\lambda\rightarrow i\rightarrow
f$, $\Gamma_{\lambda}$ and $\Gamma_i$ are the total widths of the
decaying states $\lambda$ and $i$, respectively. The sum intensity
$I_{\gamma\gamma}$ of the cascades is related to an unknown number of
intermediate levels $n_{\lambda i}=\rho\times\Delta E$ and unknown
widths of primary and secondary transitions via the equation
\begin{equation}
I_{\gamma\gamma}=\sum_{\lambda ,f}\sum_{i}\frac{\Gamma_{\lambda i}}
{\Gamma_{\lambda}}\frac{\Gamma_{if}}{\Gamma_i}=\sum_{\lambda ,f}
\frac{\Gamma_{\lambda i}}{<\Gamma_{\lambda i}> m_{\lambda i}}
n_{\lambda i}\frac{\Gamma_{if}}{<\Gamma_{if}> m_{if}} \label{eq:three}
\end{equation}

The summation is over a certain set of quantum numbers of intermediate,
initial, and final states for the purpose of comparison with the
experimental data. The thermal neutron capture cross-section for two
possible spins of compound states are listed in \cite{Mugh84}, for
example. The $J^{\pi}$ values for the initial and final cascade levels
are also known. The latter, however, is true if the energy $E_f$ of
the final state does not exceed $\simeq 1$ MeV. The optimal width of
the interval $\Delta E$ and the number $N$ of such intervals in eq.
(\ref{eq:three}) are determined by the statistics of $\gamma -\gamma$
coincidences (as a square detector efficiency) and the necessity to
obtain detailed energy dependence for $I_{\gamma\gamma}$. The width of
$\Delta E$ does not exceed 0.5 MeV even in the case of a 10\% efficiency
detector, however. The total radiative widths $\Gamma_{\lambda}$ of the
capturing states are also known from the corresponding experiments for
all stable nuclei [15].  The mean partial widths $<\Gamma_{\lambda i}>$,
$<\Gamma_{if}>$ and the total numbers $m_{\lambda i}$, $m_{if}$ of
levels excited by $E1$ and $M1$ transitions after the decay of the
states $\lambda$ and $i$, respectively, to be found in the analysis are
related to the total radiative widths as
\begin{eqnarray}
\Gamma_{\lambda}& =&<\Gamma_{\lambda i}>\times m_{\lambda i}\nonumber\\
\Gamma_i & =&<\Gamma_{if}>\times m_{if}
\label{eq:four}
\end{eqnarray}
The contribution of higher multipolarities to eqs. (\ref{eq:three}) and
(\ref{eq:four}) is smaller than the error of the determination of
$I_{\gamma\gamma}$. Equations (\ref{eq:three}) and (\ref{eq:four}) and
their obvious combination
\begin{equation}
\Gamma_{\lambda}\times I_{\gamma\gamma}=\sum_{J,\pi}\Gamma_{\lambda i}
\times n_i\times (\Gamma_{if}/<\Gamma_{if}>m_{if})
\label{eq:five}
\end{equation}
allow three ways of the estimation of the parameters of the cascade
$\gamma$-decay using the experimental data on $I_{\gamma\gamma}$ and
$\Gamma_{\lambda}$:
\begin{enumerate}
\begin{enumerate}
\item the level density can be estimated from eq. (\ref{eq:three}) using
model calculated partial radiative widths;
\item the partial widths of cascade transitions can be estimated from
eq. (\ref{eq:five}) using model calculated level densities with certain
$J^{\pi}$;
\item simultaneous estimation of the intervals of probable level
densities and radiative strength functions which satisfy eqs.
(\ref{eq:three}) and (\ref{eq:four})  in general.
\end{enumerate}
\end{enumerate}
It is clear that the level density and strength functions found
according to variants (a) and (b) inevitably contain errors caused by
the uncertainties of experimental and model values used as parameters of
the analysis. However, the influence of these uncertainties on the final
result is suppressed because of the correlation (determinated by the
used type of the functional relations (\ref{eq:three}) and
(\ref{eq:five})) between the experimental $\Gamma^{exp}_{\lambda}$,
$I^{exp}_{\gamma\gamma}$ and the parameters under study $\rho$,
$\Gamma$.

In accordance with the variant (a) the sufficiently narrow interval of
probable $\rho$ was determined for almost 30 nuclei from the mass region
$114\leq A\leq 200$ for some set of possible models of
$\gamma$-transition strength functions. An important conclusion made in
\cite{Khit98} is that the best description of the level density in the
interval from $\sim 0.5B_n$ to $B_n$ was achieved in the framework of
the generalized model of the superfluid nucleus \cite{Ign75}. Besides,
simple enough models \cite{Axel62,Kadm83} of radiative strength
functions cannot provide a correct description of the experiment and
also need modification. An analysis by variant (b) was performed by us,
as well. The main result is that there are no strength function models
for $E1$ and $M1$ transitions in deformed nuclei which could reproduce
the dependence $\Gamma_{\lambda} \times I_{\gamma\gamma}$ at primary
transition energies $E_1\leq 2-3$ MeV if the level density is set by the
model of a non-interacting Fermi-gas. Therefore, the understanding and
correct description of the $\gamma$-decay of the compound nucleus with
a high level density require experimental determination of the level
density and radiative strength functions over the entire excitation
energy region.

Further investigations \cite{Sukh99} have shown that the level density
at excitations from 1-2 to 3-4 MeV in, first of all, deformed nuclei
deviates strongly from the exponential energy dependence derived on the
basis of the idea that the nucleus is a non-interacting Fermi-gas
\cite{Dilg73}. Moreover, it is not excluded that the level density in
this energy interval can be almost constant or even decrease with
increasing excitation energy. These confirm and complement the results
obtained in \cite{Khit98}.

\section{Analysis}
The variant (c) of analysis of the experimental intensities of two-step
$\gamma$-cascades between the capturing state and several low-lying
levels allowed us to suggest an original method for the solution
(although partial) of this problem. It is based on an obvious
circumstance that $N+1$ equations (\ref{eq:three})  and (\ref{eq:four})
together with $6N$ conditions
\begin{eqnarray}
\rho(\pi=+)>0;~~~~\rho(\pi=-)>0   & &\nonumber\\
\Gamma(E1)>0;~~~~~~~\Gamma(M1)>0& &
\label{eq:six}
\end{eqnarray}
(separately for primary and secondary transitions in the case of
radiative widths) restrict some interval of possible level densities and
partial radiative widths which provide a simultaneous reproduction of
$\Gamma^{exp}_{\lambda}$ and $I^{exp}_{\gamma\gamma}$. This interval can
be estimated using modern computers and the existing computational
algorithms. Its width, however, cannot equal zero even at zero
uncertainty of the experiment. It should be added that
$I_{\gamma\gamma}$ in the form of eq. (\ref{eq:three}) is inversely
proportional (qualitatively) to the total number of states excited in
the process under study and is proportional to the ratio of cascade
transition widths to their mean values. Therefore, the method of
analysis described below has a maximum sensitivity at minimum density
of the excited states (unlike the methods \cite{Svir88,Ign78}).

As in the case of other reactions (followed by $\gamma$-emission) used
for the determination of $\rho$, all values obtained experimentally in
the $(n_{th},\gamma )$ measurements are determined by the product
$\Gamma_{\lambda i}\times\rho$. Hence, in the calculation deviation of
one of the two parameters from its mean value is compensated by
deviation of the other one with the corresponding magnitude and sign.
This circumstance should be taken into account in data processing --- a
 minimum or maximum value of the level density derived from the
experimental data results, e.g., in a maximum or minimum value of the
corresponding strength functions.

It should be noted that deviation of the calculated level density from
the true value is completely compensated by deviation of strength
functions when $\Gamma_{\lambda}$ is only calculated. In the case of the
calculation of $I_{\gamma\gamma}$ the compensation is incomplete. This
very circumstance allows one to select the intervals of $\rho$ and
$\Gamma_{\lambda i}$ which provide the description of the
$I_{\gamma\gamma}$ and $\Gamma_{\lambda}$ parameters with an acceptable
uncertainty. This analysis can be performed by means of finding large
enough sets of random  values of $\rho$ and $\Gamma_{\lambda i}$ which
reproduce completely the parameters $\Gamma^{exp}_{\lambda}$ and
$I^{exp}_{\gamma\gamma}$ and belong to the intervals that contain true
values. This means that most probable values of the level density  and
radiative strength functions of dipole $\gamma$-transitions and
intervals of their uncertainties can be found by selection of pairs of
random $\rho$ and $k$ which satisfy, in general, eqs.
(\ref{eq:three}) and (\ref{eq:four}) or (\ref{eq:three}) and
(\ref{eq:five}). This requires numerous repetitions of the procedure
and statistical methods of analysis.

It is clear that the widths of the intervals of probable $\rho$ and
$k$ satisfying eqs. (\ref{eq:three}) and (\ref{eq:four}) increase
with increasing number of unknown parameters in the equations. According
to experimental conditions, the summation in eqs. (\ref{eq:three}) and
(\ref{eq:five}) as over all intermediate states of the cascades. Since
the summed data included cascade transitions of different
multipolarities, we could not obtain the strength functions of $E1$ and
$M1$ transitions and the level density for different parities separately
with a good precision. In practice, from a combination of eqs.
(\ref{eq:three}) and (\ref{eq:five}) the sum of strength functions and
the sum of level densities of both parities should be only derived and
compared with model predictions. The corresponding summation reduces
considerably the uncertainty of the observed result due to
anti-correlation of elements.

Indeed, an analysis of the available data confirms that the dispersion
of each set of $\rho(\pi=+)$, $\rho(\pi=-)$, $k(E1)$ and $k(M1)$
random values is too large to make any conclusions about independent
correspondence of individual values to the model.

A sufficiently large $N$ and the nonlinearity of eqs. (\ref{eq:three})
and (\ref{eq:four}) stipulate the choice of the way to solve the system
of equations and inequalities - the Monte Carlo method. The simplest
iterative algorithm \cite{Khit98} was used for this aim: we set some
initial values for $\Gamma(E1)$, $\Gamma(M1)$, $\rho(\pi=-)$, and
$\rho(\pi=+)$ and then distort them by means of random functions. If
these distortions  decrease the parameters $\Delta=(I_{\gamma\gamma}^
{exp}-I_{\gamma\gamma}^{cal})^2$  at this step of the iteration
procedure, then the distorted values are used as initial parameters for
the next iteration. Agreement between the experimental and calculated
cascade intensities and the total radiative widths, respectively, is
usually achieved after several thousand iterations. As a result we get
two random ensembles of level densities and partial widths for every $N$
energy intervals. Examples of intermediate and final results of one of
many variants of the calculation for two nuclei are shown in Figs.~1 and
2. It is obvious that such iterative process can be realized in an
unlimited number of ways. We chose a sufficiently simple and effective
way: the Gaussian curve is used as a distorting function for logarithms
of $\rho$ and $f$
\begin{equation}
f(E)=A \times\exp (-(E-E_0)^2/{\sigma}^2)
\label{eq:seven}
\end{equation}
Its parameters are independently chosen for the level density and
strength functions from the intervals [-0.2;0.2], $[E_d;B_n]$ and
[0.3 MeV;$B_n]$ for $A$, $E_0$, and $\sigma$, respectively using a
standardized random value distributed uniformly in [0;1]. Here $E_d$ is
the maximum excitation energy of the known discrete level involved in
the calculation. Numerous repetitions of the iterative calculation with
different initial parameters (including obviously unreal values of
$\Gamma$ and $\rho$) for $\sim$30 nuclei from the mass region $114\leq
A\leq 200$ show that this algorithm yields rather narrow intervals of
the sum level density of both parities and of the sum partial widths of
$E1$ and $M1$ transitions. The use of eq. (\ref{eq:seven}) with mentioned
parameters allows one to get a set of different, smooth enough
functional dependences for both $\rho$ and $k$. In this case for the
majority of the studied nuclei the values of level density are in good
agreement with the number of the observed intermediate levels of the
cascades resolved as the pairs of peaks. In some nuclei, however, the
mean level density  (which together with the mean strength functions
provides reproduction of cascade intensities) is less than the number
of intermediate levels observed below  $\approx 2$ MeV. The main portion
of this discrepancy is removed in all cases if one foresees a
possibility of additional local variation of $k$ for high-energy
transitions in the energy interval which, as a rule, does not exceed
0.1-0.2 MeV. One of the examples of this kind is shown in Fig.~1. The
necessity to account this effect can be due to both insufficient
averaging of the random partial widths of primary transitions and their
possible dependence on the structure of the excited low-lying level.
This can result, for instance, from concentration of the strength
of the fragmented single-particle or phonon states.

\section{Asymptotical uncertainty of the obtained parameters}
The method suggested by us for estimation of
$\rho$ ¨ $k$ cannot give unique value of these parameters at a given
energy of the excitation or quantum energy. Therefore the question arises
about the value of their uncertainty at different energies and degree of
possible systematical deviations of the observed parameters
from the modal values. The results of modelling for $^{156}Gd$ and $^{198}Au$
shown in Fig.~3 answer these questions. Intensity of cascades for these
nuclei were calculated under assumption that the strength function $k(E1)$
is described by model \cite{Axel62} and value of $k(M1)=const$;
level density exponentially increases with the energy or have some step-like
structure. Below the excitation energy $\approx 1-2$ MeV the calculation
used experimental decay scheme.Consequently, the calculated intensity
distribution of cascades in function of the primary transition energy
has one or two maxima. (Other conditions of the calculation completely
corresponded to the experiment).

Figure 3(b) shows that the model level density is reproduced practically
without systematical error and the width of the interval of its probable
values does not exceed 20-30\%.

Discrepancy between the experimental and model sum $k(E1)+k(M1)$
results from that the total radiative width calculated according model
\cite{Axel62} does not correspond to the experimental
value. Energy dependence of $k(E1)+k(M1)$ is reproduce rather well -- sharp
changes in the first derivative with respect to the quantum energy is not
observed (unlike some other nuclei studied by us).

So, one can summarize that the suggested method provides reliable enough
estimation of the level density and radiative strength functions of dipole
transitions.

\section{Approach used in calculation}
The insufficient experimental data on cascade $\gamma$-transitions (only
cascades terminating at low-lying levels ($E_f<1$ MeV) of nuclei were
studied \cite{Bon99}) does not allow us to determine the level densities
and gamma-widths without the following important assumption: the
strength functions of transitions of a given multipolarity only depend
on the transition energy and do not depend on the structure and energy
of the corresponding excited states. Their nonequal values for
$\gamma$-transitions of equal energies but populating different levels
is, in part, compensated by the circumstance that the left part of eq.
(\ref{eq:five}) depends on absolute radiative strength function values
of primary transitions and depend only on the ratio of strength
functions in the case of secondary transitions. These decrease the
effect of the discussed assumption on the $k(E1)+k(M1)$ values but do
not remove it  completely. There is no necessity in introduction of any
hypotheses of spin dependence of level density differing from that
predicted in models \cite{Dilg73,Ign75}.

\section{Sources of errors in the determination of strength functions
and level densities}
The presence of the statistic and systematic errors in determination of
$I_{\gamma\gamma}$, $\Gamma_{\lambda}$ and specific problems of
extraction of level density and radiative strength functions cause
noticeable uncertainties of the determined parameters. The influence of
the different sources of errors on the obtained results manifest itself
in a different degree.
\begin{enumerate}
\item Uncertainties of the measuring of terms in eqs. (\ref{eq:three})
and (\ref{eq:five}) result in errors of strength functions and level
density. Owing to a linear relation between $\Gamma_{\lambda}$,
$I_{\gamma\gamma}$ and $\Gamma_{\lambda i}$ in eq. (\ref{eq:five}),
$\simeq 10\%$ errors of $\Gamma_{\lambda}$ and $I_{\gamma\gamma}$
achieved in the experiment cause rather a small error in the
determination of $\Gamma_{\lambda i}$ and $\rho$ as compared to
dispersion of the obtained data.
\item The more considerable source of uncertainty in the determination
of the strength functions and $\rho$ is a systematic error of
decomposition \cite{Bon391,Bon95} of the experimental spectra into two
components corresponding to solely primary and solely secondary
transitions. The analysis \cite{Sukh99} showed that the error in
$\Delta I_{\gamma\gamma}$ caused by this procedure does not usually
exceed $\simeq 20\%$ for primary  transition energy $E_1<3-4$ MeV.
Intensities of cascades (histograms in Figs.~1, 2, 4-13) at these primary
transition energies can be overestimated, as a maximum, by the above
value. At the higher energies they can be decreased by the same value
(the total intensity is preserved). In order to estimate the influence
of $\Delta I_{\gamma\gamma}$ on the final results, the
$I_{\gamma\gamma}$ values were varied within a level of 25\%. These
variations caused changes in $k(E1)+k(M1)$ and $\rho$ which did not
exceed the dispersion of the data plotted in Figs.~4-13.
\item The maximum uncertainty of level density and radiative strength
functions results from the use of condition (6). It dominates at any
possible precision in determination of $I_{\gamma\gamma}$ and
$\Gamma_{\lambda}$. The simplest way to estimate these errors at any
$E_1$ and $E_{ex}$ is the following:
\begin{enumerate}
\item taking into account that the probabilities of deviations with
opposite sign of the random $\rho_i$ and  $f_i$ values with respect to
their mean values are equal and decrease as the absolute values of
deviations increase; and
\item assuming that mathematical expectations of the random ensembles of
the $\rho_i$ and  $k_i$ values satisfying eqs.
(\ref{eq:three})-(\ref{eq:five}) correspond to their real values
\end{enumerate}
one can consider the mean-square deviations of the random values
relative to their arithmetical means as the estimations of the errors.
These errors can be attributed to level density and strength functions
separately in spite of their strong anti-correlation. Just these
uncertainties are shown for the radiative strength functions and level
density plotted in Figs.~4-13.

On the whole we can summarize the situation as the following. At the presently
achieved accuracy for experimental determination of $I_{\gamma\gamma}$
and $\Gamma _{\lambda}$, level densities and strength functions are
derived from eqs. (\ref{eq:three})-(\ref{eq:five}) with the mean total
uncertainties of about 40-50\% in the worst case. Asymptotic value of
this uncertainty at zero statistic and systematic errors of the
experiment is equal, in the average, for both $\rho$ and $k$ and cannot
be less than $\simeq$ 20\%.
\item There are two ways to decrease the errors of the level density
and strength functions determined from eqs.
(\ref{eq:three})-(\ref{eq:five}):
\begin{enumerate}
\item the increase of the volume of the experimental data on the cascade
intensities;
\item the reduction of the number of parameters in eqs.
(\ref{eq:three})-(\ref{eq:five}) owing to the use of additional
information or introduction of some new assumptions.
\end{enumerate}
In the first case the problem can be easily solved experimentally: the
use of a Compton-suppressing spectrometer consisting of $HPGe$ detectors
with an efficiency of not less than 30-40\% allows the selection from a
mass of $\gamma -\gamma$ coincidences of two-step cascades for a
considerably larger number of their final levels than at present. From a
combination of eq. (\ref{eq:three})  for the sum over all final levels
of cascades and an individual final level $f$ one can determine
the ratio $\Gamma_{if}/<\Gamma_{if}>\times m_{if}$ for all possible
values of $i$ and $f$, i. e., determine energy dependence of the
experimental sum $k(E1)+k(M1)$ for any possible secondary transitions,
get rid of the only approach used in the analysis, and reduce the number
of parameters in the analysis.
\end{enumerate}
The data shown in Figs.~4-13 were obtained under assumption about a
constancy of the ratio
\begin{equation}
\Gamma_{\lambda i}/\Gamma_{if}=const
\label{eq:eight}
\end{equation}
for the transitions with equal multipolarity and energy $E_1$ in all
interval of the neutron binding energy.

The comparison of the total $\gamma$-spectra and population of low-lying
levels calculated in this way with the available experimental data
including spectroscopic information \cite{Bon99} shows that even such
assumption provides better accuracy in calculating the parameters of
cascade $\gamma$-decay than the approach using the models
\cite{Dilg73,Ign75,Axel62,Kadm83}. Unfortunately, we could not achieve
complete correspondence between the estimated level density and
available spectroscopic information. Nevertheless, the obtained values
demonstrate certain correspondence of our level density with the numbers
of the excited levels observed in the experiment \cite{Bon99}. Some nuclei,
however, demonstrate residual discrepancy (for example, $^{170}Tm$,
Fig.~1). This discrepancy can be attributed, partially, to both
insufficient precision of assumption (\ref{eq:eight}) and inexactitude
of the spectroscopic data. Their errors can be also considerably
decreased using more efficient spectrometer of $\gamma -\gamma$
coincidences than that used by authors \cite{Bon99}.

On the whole, in spite of the uncertainties mentioned above one can
conclude that at a given stage of the experimental investigation of the
cascade $\gamma$-decay of compound states our method provides more
reliable results than methods \cite{Svir88,Ign78,Bart73}.

\section{Main results of analysis}
The type of relation between $k$ and $\rho$ on the one hand and
between $\Gamma_{\lambda}$ and $I_{\gamma\gamma}$ on the other hand
does not allow one to determine $k$ and $\rho$ unambiguously and
independently. Some deviation of, for example, $\rho$ from a real value
is inevitably compensated by deviation of strength functions of the
corresponding magnitude and sign. Nevertheless, the results obtained
in the present analysis can be used for the verification of nuclear
models and, if necessary, for the determination of the direction of the
further development of these models. The main argument in favour of this
statement is relatively week dependence of the final results on the
initial values of strength functions  and $\rho$ in the iterative
process. As an example, Figs.~1 and 2 show the strength function and
$\rho$ values obtained for their unreal initial values: $\rho (E_{ex})=
\rho (B_n)$, the strength functions decrease linearly as the
transition energy increases. Nevertheless, the final results of the
iterative process quite agree with a general picture obtained for a
large enough set of different real and unreal initial values of $k$
and $\rho$. This confirmes the conclusion that the strength functions
and level density obtained from the analysis can be considered as most
probable.

The strength functions $k(E1)+k(M1)$ and level densities $\rho$ obtained
in the present analysis are plotted in Figs.~4-13. For
every set of random $\rho$ at a given excitation energy $E_{ex}$ and
$k(E1)+k(M1)$ at a given primary transition energy $E_1=B_n-E_{ex}$
there were determined both their mean values and probable dispersion
using usual relationships of statistical mathematics. The results of the
analysis are compared with predictions of the level density models
\cite{Dilg73,Ign75} and models of radiative widths \cite{Axel62,Kadm83}.
In the case of radiative strength functions a comparison is performed in
the following manner: the $k(E1)$ values calculated according to the
models \cite{Axel62} and \cite{Kadm83} (upper and lower curves,
respectively) are summed with $k(M1)=const$ which is normalized so that
the ratio $\Gamma(M1)/\Gamma(E1)$ would be approximately equal to the
experimental data at $E_{\gamma}\simeq B_n$.

A comparison of the results of the analysis with predictions of the
models \cite{Dilg73,Ign75,Axel62,Kadm83} (often used by
experimentalists) shows that:
\begin{enumerate}
\item the sums $k(E1)+k(M1)$ and $\rho$ are not monotonic functions of
the energy and, probably, reflect the most common peculiarities of the
structures of the states connected by the corresponding
$\gamma$-transitions;
\item the energy dependence of $k(E1)+k(M1)$ differs strongly from
predictions of the models \cite{Axel62,Kadm83} in the case of even-even
compound nuclei from the region of the $4s$-resonance of the neutron
strength function, at least;
\item the $k(E1)+k(M1)$ functions increase from near-magic to deformed
nuclei and from complicated highly-excited states to simpler low-lying
levels which are populated by $\gamma$-transitions under consideration;
\item relative deviations of the obtained strength functions and level
densities from the mean values are characterized by strong negative
correlation. In the majority of nuclei the correlation coefficient
changes from -0.6 to -1.0. This means that the strength functions and
level densities are not independent variables in eqs. (\ref{eq:three})
and (\ref{eq:five}), which provides the possibility of their
simultaneous determination;
\item the probable level density determined in the present analysis
conforms to the picture obtained in previous experiments
\cite{Khit98,Sukh99}: up to  the excitation energy 1-2 MeV, our data
are not in contradiction with the exponential extrapolation of
$\rho (E_{ex})$ predicted by the Fermi-gas back-shift model
\cite{Dilg73}. The energy dependence of the level density in the
interval from 1-2 to some threshold value $E_b$ is considerably weaker
than it follows from any existing level density model. Above
$E_b\approx 3$ MeV for $N$-odd and $\approx 4$ MeV for $N$-even nuclei,
the level density, most probably, corresponds better to the predictions
of the generalized model of the superfluid nucleus in its simplest form
\cite{Ign75}.
\end{enumerate}
This change in the behaviour of the level density in the vicinity of the
excitation energy $E_b$ may signify a qualitative change in the nuclear
properties. The observation \cite{Sukh97} of the probable harmonicity of
the excitation spectra of the intermediate levels of the most intense
cascades in a large group of nuclei from the mass region $114\leq A\leq
200$ allows an assumption that the nuclear properties at low energy are
mainly determined by vibrational excitations (probably, a few phonons of
rather high energy). A very quick exponential increase in the level
density above $E_b$ says about the probable dominant influence of the
inner, many-quasiparticle type of excitations of these states.

\section{Discussion}
The method suggested in present work allows model independent,
simultaneous estimation of intervals of probable values of the level
densities with given spins and summed strength functions of primary
dipole transitions populating them. The method is effective in
investigations of any stable nucleus. The main
differences of this algorithm from the known methods of determination
of level densities \cite{Svir88,Ign78,Vona83} and radiative strength
functions \cite{Bart73} are the following:
\begin{enumerate}
\item Our method does not permit one to get the sole values of $\rho$
and $k$ for a given energy. But the width of the intervals of their
probable magnitudes depends very weakly on the uncertainty in
determination of $\Gamma_{\lambda}$ and $I_{\gamma\gamma}$ at the
achieved precision of the experiment, at the one hand, and is narrow
enough in order to get new information on nuclear matter, from the other
hand.
\item The most correct and reliable data on the level density is derived
from the evaporation spectra at the highest excitation energies;
analysis of the cascade intensities provides similar data for the lowest
energies. So, both methods mutually add each other.
\item Analysis of cascade intensities allows direct determination of the
absolute level densities, evaporation spectra usually provide
\cite{Svir88,Ign78} information on nuclear temperature.
\item Systematical uncertainties of both methods do not relate.
Discrepancies in the independently determined level densities at some
energies indicate to necessity, for example, to determine more precisely
the barrier transmission factor for the evaporated particle or to take
into account different energy dependence of $k$ of the primary and
secondary transitions of the $\gamma$-cascades. Besides, they can
testify to necessity to describe more correctly direct and
pre-equilibrium processes in nuclear reactions for deformed nuclei or to
define more precisely the nuclear excitation energy above which
thermodynamical parameters of a nucleus are determined mainly by
quasiparticle excitations.
\item Energy dependence of the data in Figs.~4-13 can be reproduced well
enough in the framework of modern version of the generalized model of
the superfluid nucleus \cite{Ign85} if the temperature of the phase
transition is diminished up to the value $T'_{cr}\approx 0.7T_{cr}$,
where
\begin{equation}
T_{cr}=\delta/1.76
\label{eq:nine}
\end{equation}
is the temperature of the transition from the superfluid to normal phase
of homogeneous Fermi-system \cite{Bard57}. But re-determination of the
entropy and temperature predicted by model \cite{Ign85} should be done
so that nuclear temperature below $T'_{cr}$ will not increase with
decreasing excitation energy.
\item Additional and independent arguments in favor of reliability of
step-like structure in level density are:
\begin{enumerate}
\item combinatorical calculation \cite{Vdov76} of density of the states
with $K^{\pi}=1/2+$ in  $^{165}Dy$ below $B_n$, providing similar to
Figs.~4-13 picture;
\item analysis \cite{Zhur99} of the experimental data from the reaction
$^{165}Ho(p,n)^{165}Er$. This also demonstrates some step-like structure in
the total level density at low excitations;
\item precise analysis \cite{Masl00} of the neutron cross sections for
actinides testify to necessity to take into account the influence of
the pairing interaction on the level density for the wide interval of
the neutron energies manifesting itself, in particular, as
irregularities in the energy dependence of the level density.
\end{enumerate}
\item It is obvious that the structures shown in Figs.~4-13 can be
inherent not to the total level density with given $J^{\pi}$, but only
to that part of them which are really excited in $(n,\gamma)$ reaction.
Then, unlike the existing notions, this reaction is selective and
structures of the excited states must be taken into account in any
calculations of parameters of this reaction in the entire excitation
energy region below $B_n$.
\end{enumerate}

\section{Conclusions}
A new method is suggested for a simultaneous estimation of the probable
level density populated by dipole primary transitions in the
$(n_{th},\gamma )$ reaction and the sum strength functions $k(E1)+k(M1)$
of these transitions. Unlike other methods used for the investigations
of nuclear properties below the excitation energy 6-9 MeV, this method
allows the estimation of $\rho$, radiative strength functions, and
intervals of their probable variations without any model notions of the
nucleus.

The method is universal -- it can be used for any nucleus and reaction
with $\gamma$-emission. The latter is possible if the excitation energy
interval of high-lying states is narrow enough in order to use the sum
coincidence technique. Besides, the most probable quanta ordering in the
cascades must be determined for the main part of the observed cascade
intensity. It should be noted, that in the case of a lack of the
experimental values of the total radiative widths of decaying high-lying
states the absolute radiative strength functions cannot be determined.
In this case only relative energy dependence of the radiative strength
functions can be obtained.

The most important (although preliminary and qualitative) physical
result is that the level density below the neutron binding energy
(first of all in deformed nuclei) cannot be reproduced to a precision
achieved in the experiment without more precise than in \cite{Ign85}
accounting for the co-existence and interaction of superfluid and usual
phases of nuclear matter in this whole excitation energy interval.

The obtained results demonstrate very serious and obvious discrepancies
with the existing ideas of the structure of the deformed nuclei. These
data agree completely with an earlier obtained qualitative picture
\cite{Sukh97} of the studied process: considerable influence of
vibrational excitations on the nuclear properties below the excitation
energy $E_b$ and a transition to dominant influence of quasiparticle
excitations above this energy.

{This work was supported by RFBR Grant No. 99-02-17863}

\begin{figure}
\begin{center}
\leavevmode
\epsfxsize=12.5cm
\epsfbox{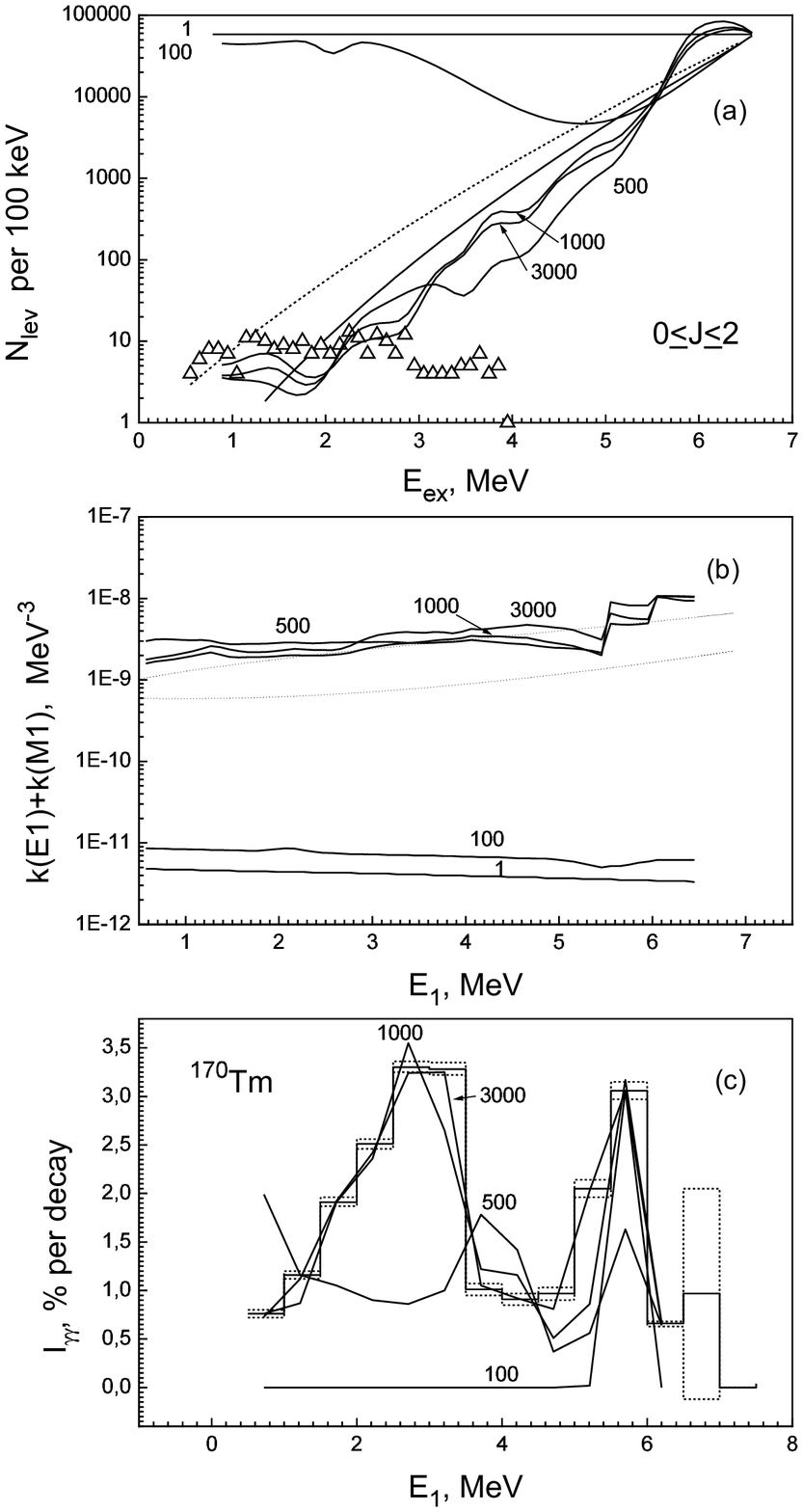}
\end{center}
\hspace{-0.8cm}

{\bf Fig.~1.}~The examples of $\rho$ (a) and $k$ (b) intermediate
values and the corresponding distributions of cascade intensities (c)
for the $^{170}Tm$ odd-odd nucleus in function of the primary transition
energy $E_1$  or excitation energy $E_{ex}$. Letters next to the lines
mean the number of iterations. Triangles show number of levels excited
by the primary dipole transitions with the energy $E_1$ in the energy
interval of 100 keV. The dashed curve (a), (b) represents model
predictions, the histograms (c) represent the experimental cascade
intensities with statistical errors.

\end{figure}
\newpage
\begin{figure}
\begin{center}
\leavevmode
\epsfxsize=12.5cm
\epsfbox{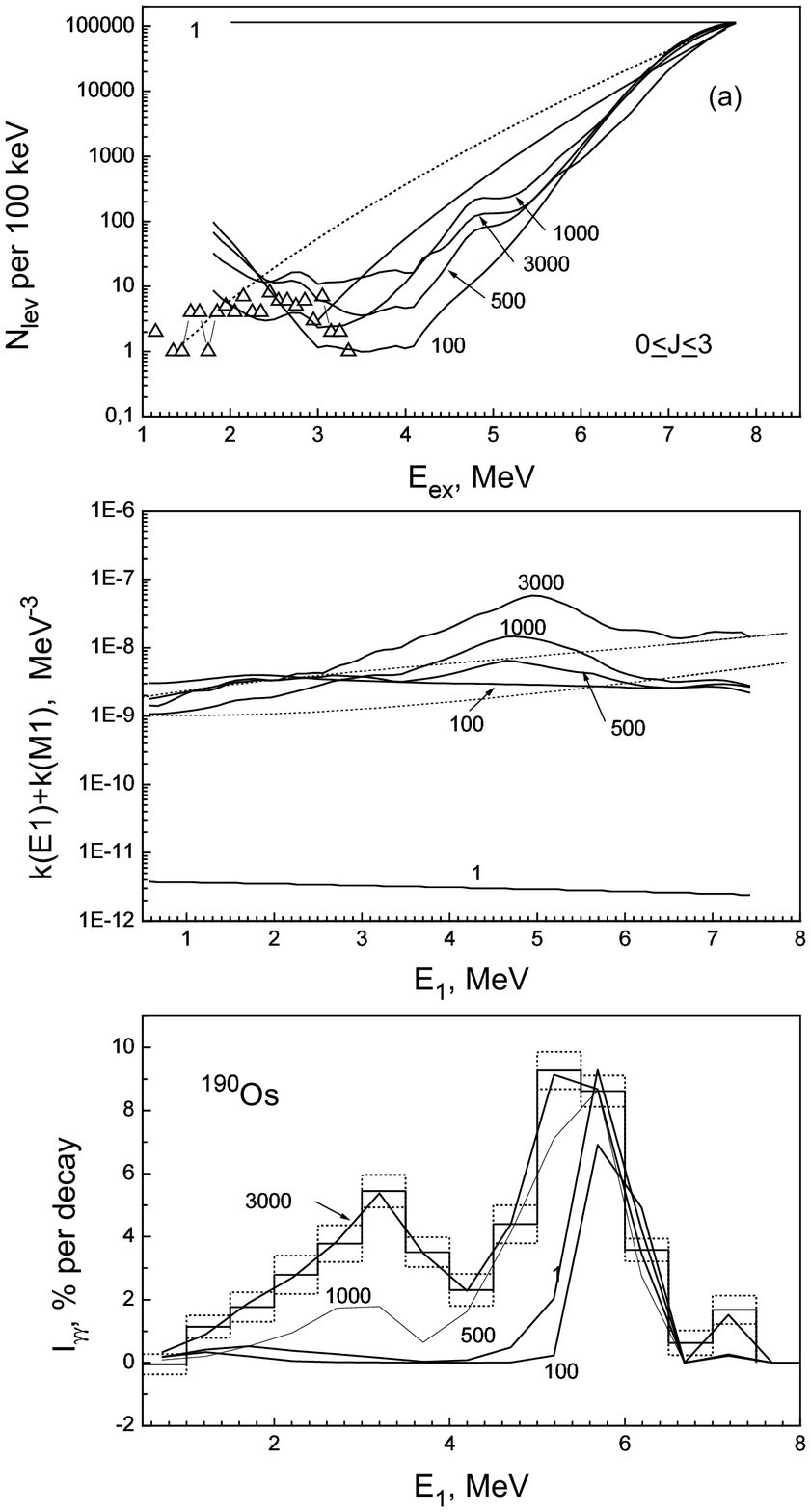}
\end{center}
\hspace{-0.8cm}

{\bf Fig.~2.}~The same as in Fig.~1,  for the $^{190}Os$ even-even
nucleus.
\end{figure}
\newpage
\begin{figure}
\begin{center}
\leavevmode
\epsfxsize=12.5cm
\epsfbox{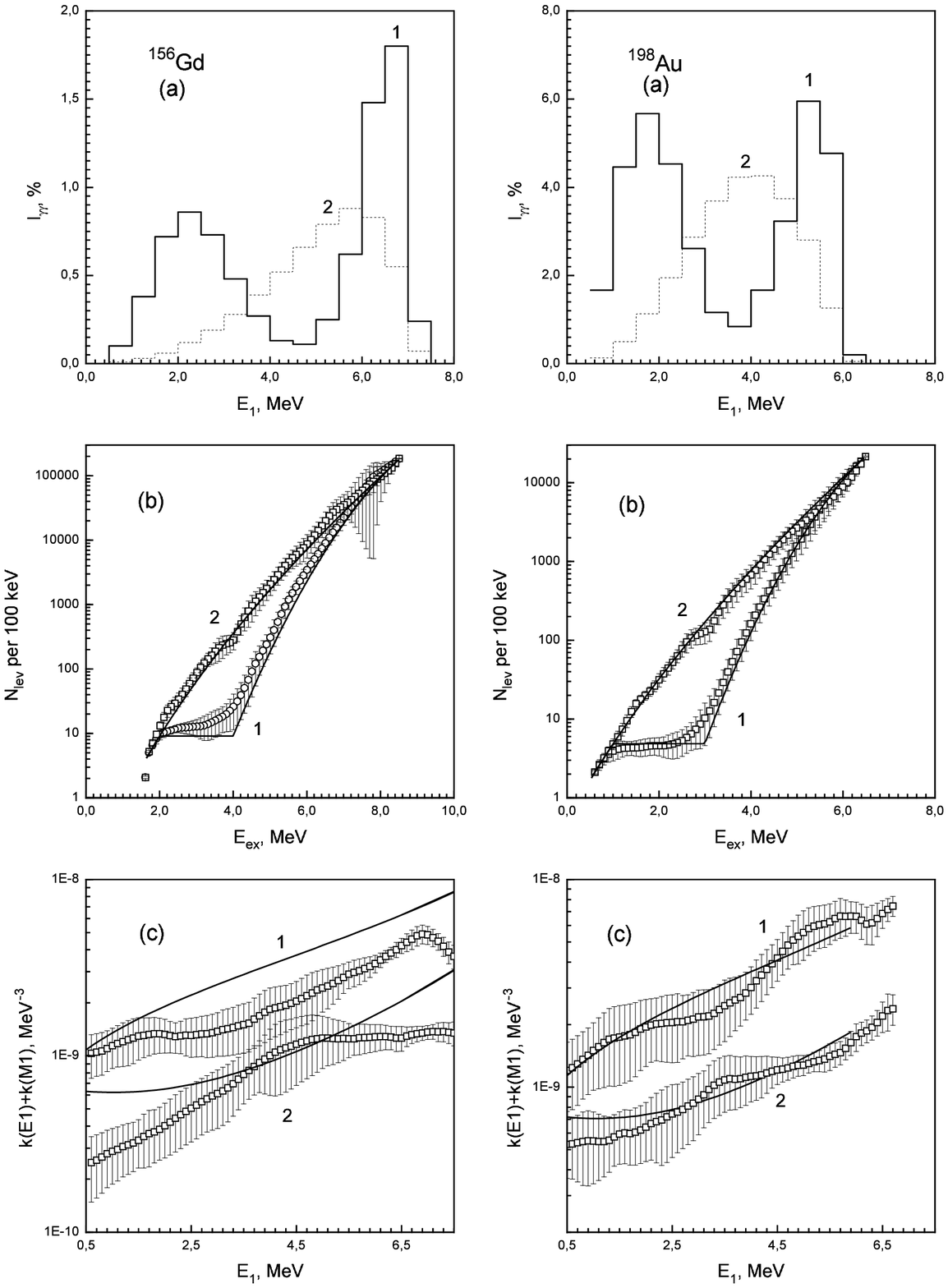}
\end{center}
\hspace{-0.8cm}
{\bf Fig.~3.}~The intensity of two-step cascades (a) calculated
using level density \cite{Dilg73,Ign75}
shown by solid lines in (b) and radiative strength \cite{Axel62,Kadm83}
functions - line 1 in (c) (line 2 in (c) represents predictions of model [5]).
Points with error bars represent the interval of possible values of $\rho$
(b) and $k$ (c) providing acceptable precision in reproduction of cascade
intensities shown in (a).

\end{figure}
\newpage
\begin{figure}
\begin{center}
\leavevmode
\epsfxsize=12.5cm
\epsfbox{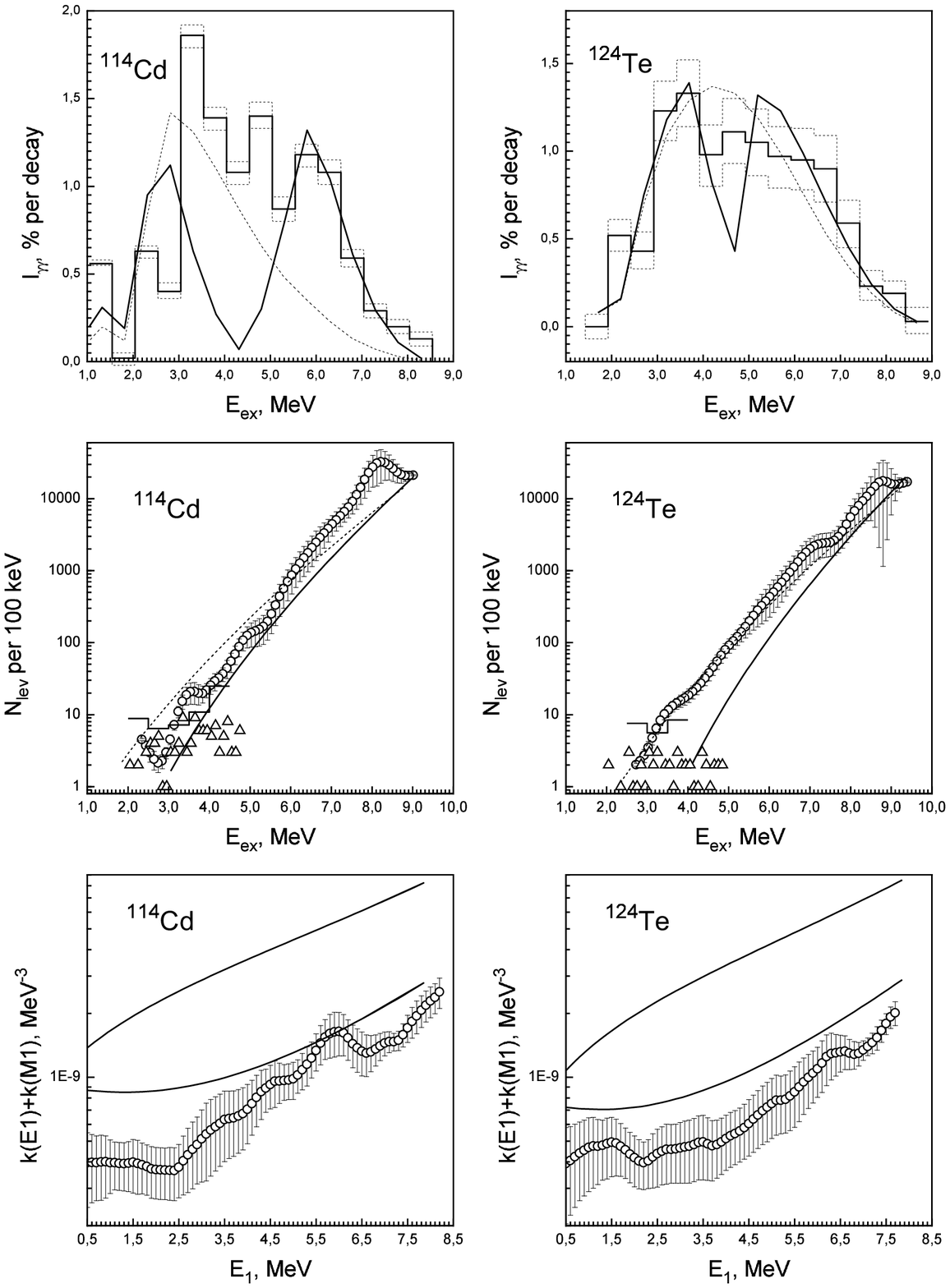}
\end{center}
\hspace{-0.8cm}

{\bf Fig.~4.}~Experimental cascade intensities $I_{\gamma\gamma}$ in 0.5 MeV
energy bins with ordinary statistical errors for $^{114}Cd$ and $^{124}Te$
(histograms. Curves represent calculation performed like
that shown in Fig.~3. Points with errors represent number of levels per 100
keV energy interval and sums $k(E1)+k(M1)$, respectively.

\end{figure}
\newpage
\begin{figure}
\begin{center}
\leavevmode
\epsfxsize=12.5cm
\epsfbox{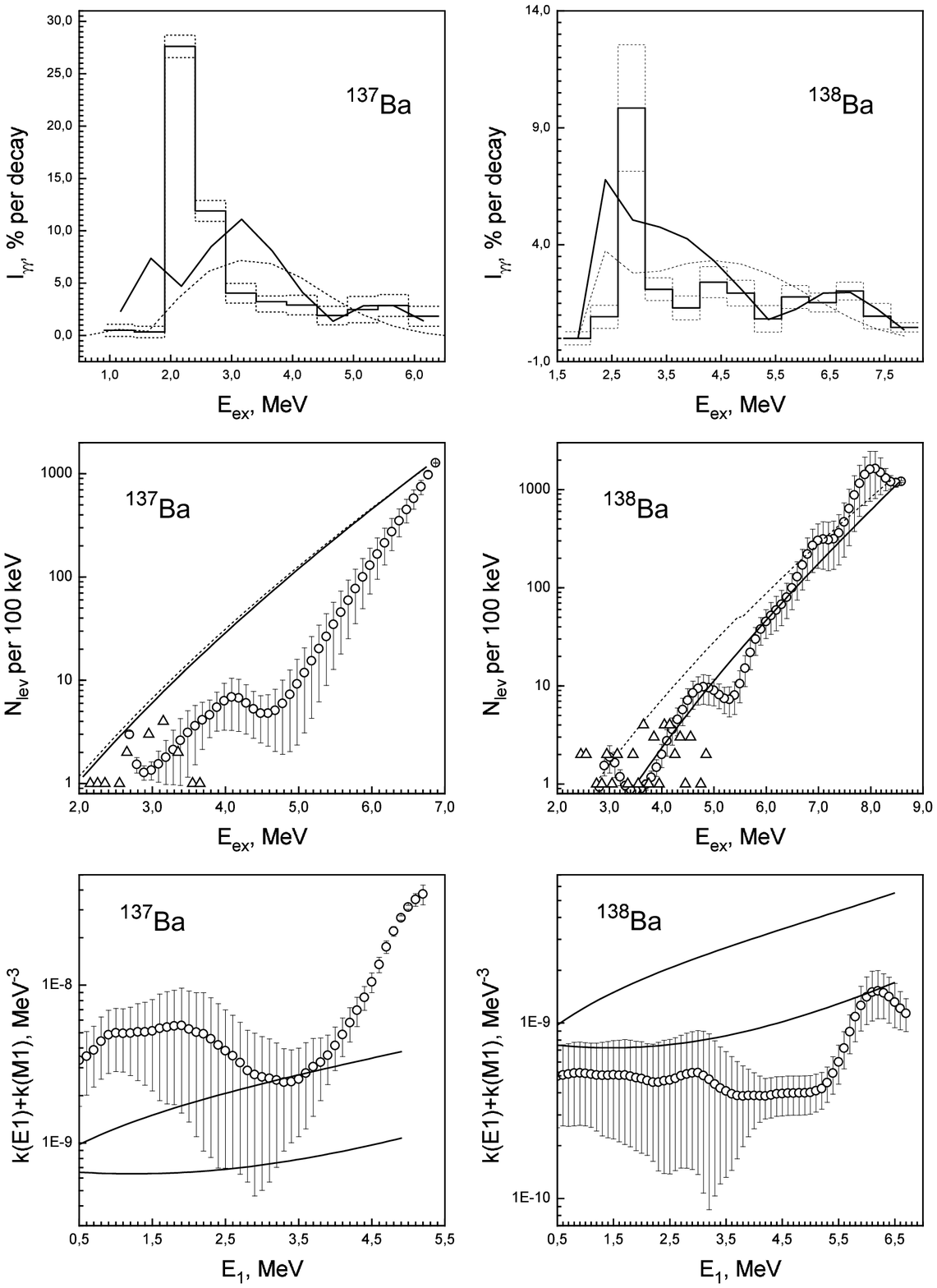}
\end{center}
\hspace{-0.8cm}

{\bf Fig.~5.}~The same as in Fig.~4 for $^{137}Ba$ and $^{138}Ba$.

\end{figure}
\newpage
\begin{figure}
\begin{center}
\leavevmode
\epsfxsize=12.5cm
\epsfbox{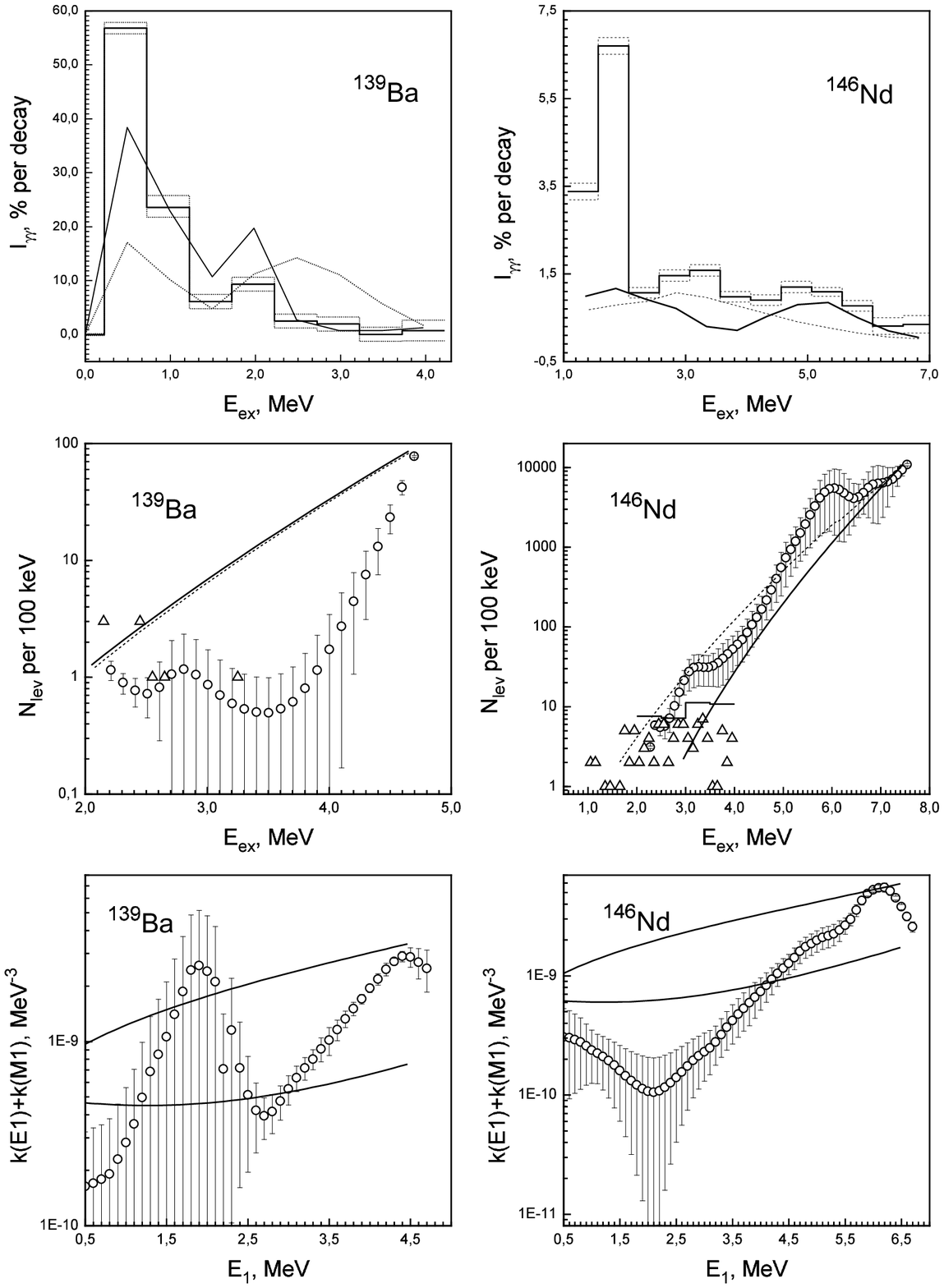}
\end{center}
\hspace{-0.8cm}

{\bf Fig.~6.}~The same as in Fig.~4 for $^{139}Ba$ and $^{146}Nd$.

\end{figure}
\newpage
\begin{figure}
\begin{center}
\leavevmode
\epsfxsize=12.5cm
\epsfbox{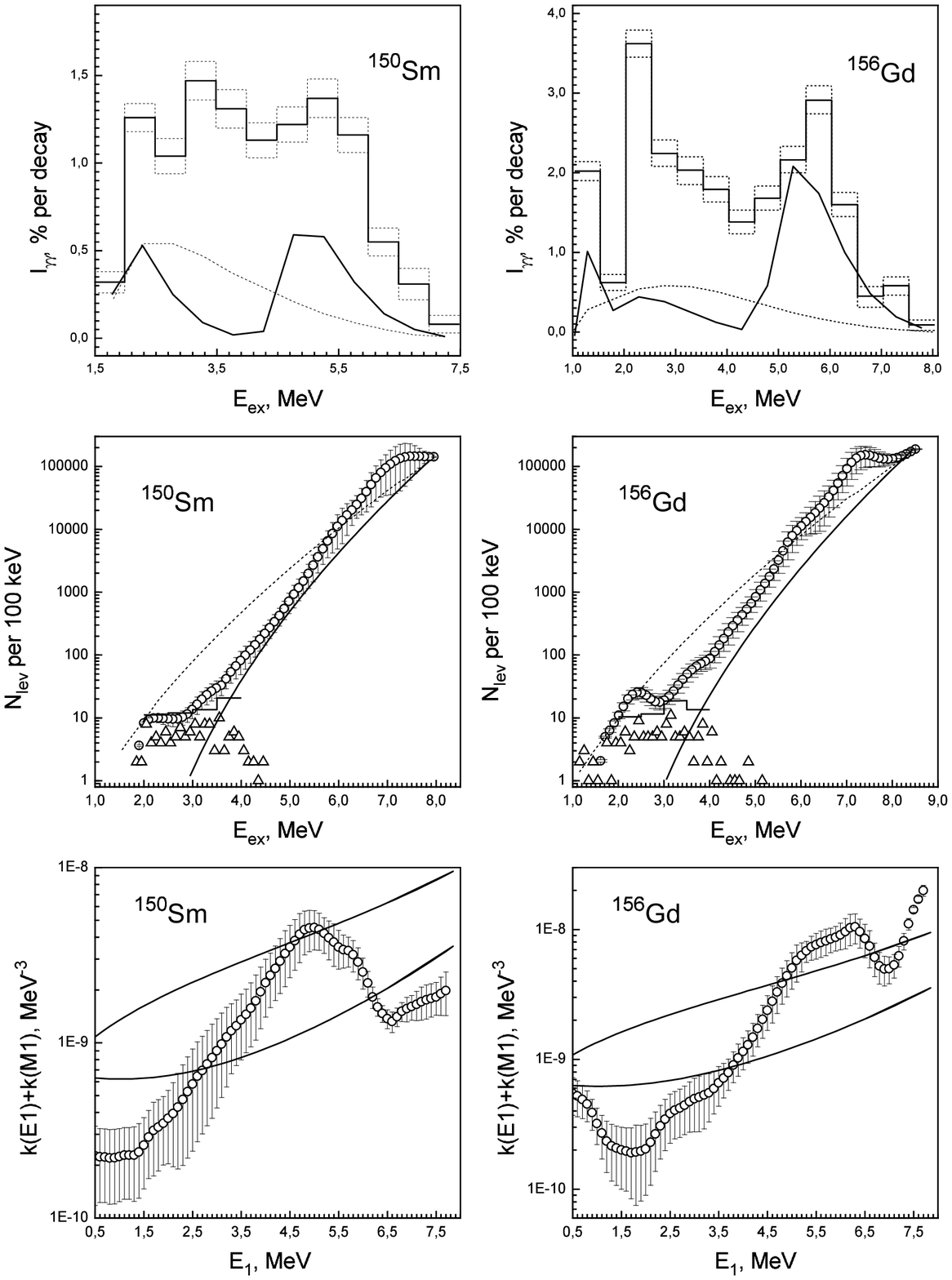}
\end{center}
\hspace{-0.8cm}

{\bf Fig.~7.}~The same as in Fig.~4 for $^{150}Sm$ and $^{156}Gd$.

\end{figure}
\newpage
\begin{figure}
\begin{center}
\leavevmode
\epsfxsize=12.5cm
\epsfbox{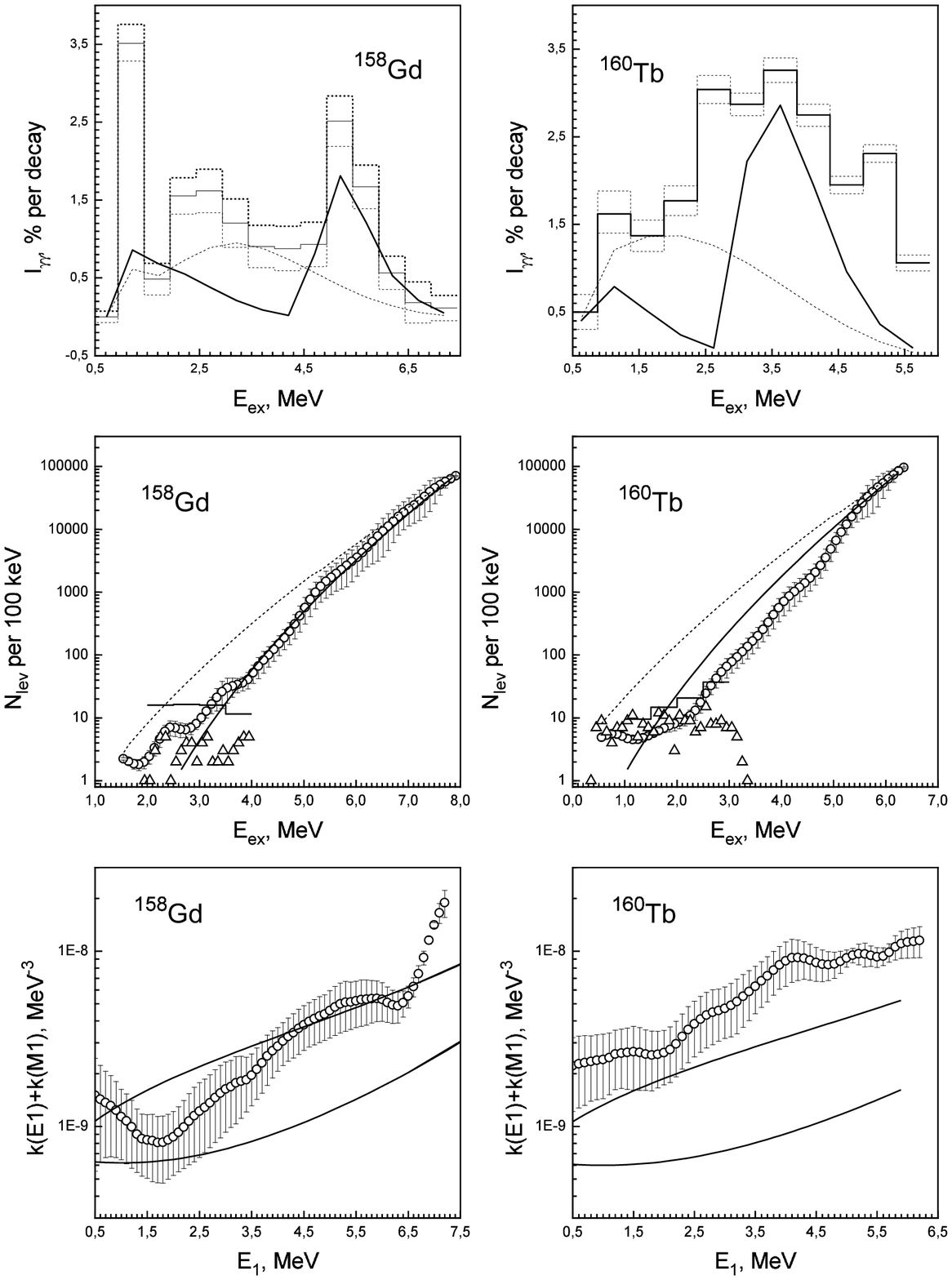}
\end{center}
\hspace{-0.8cm}

{\bf Fig.~8.}~The same as in Fig.~4 for $^{158}Gd$ and $^{160}Tb$.

\end{figure}
\newpage
\begin{figure}
\begin{center}
\leavevmode
\epsfxsize=12.5cm
\epsfbox{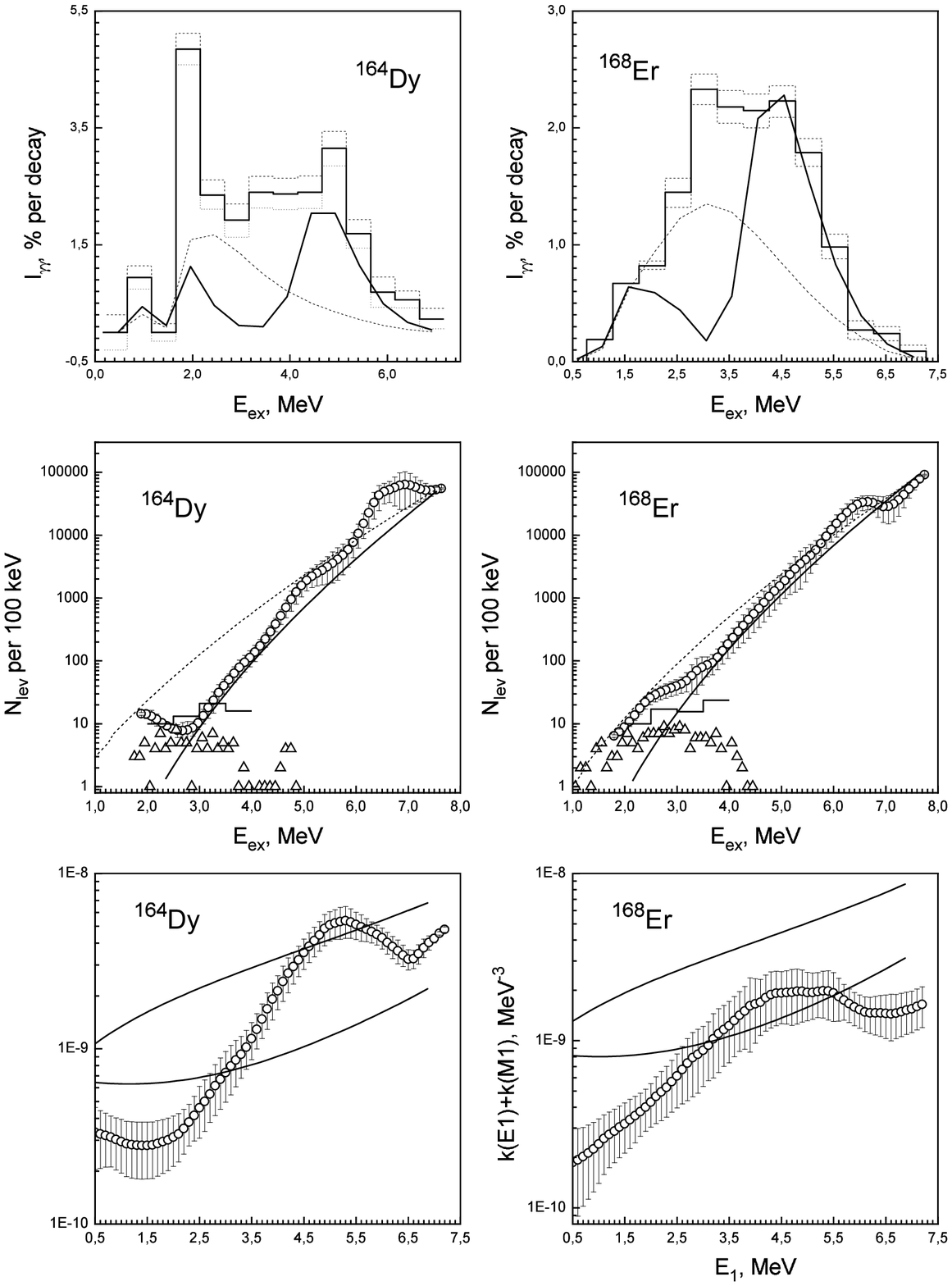}
\end{center}
\hspace{-0.8cm}

{\bf Fig.~9.}~The same as in Fig.~4 for $^{164}Dy$ and $^{168}Er$.

\end{figure}
\newpage
\begin{figure}
\begin{center}
\leavevmode
\epsfxsize=12.5cm
\epsfbox{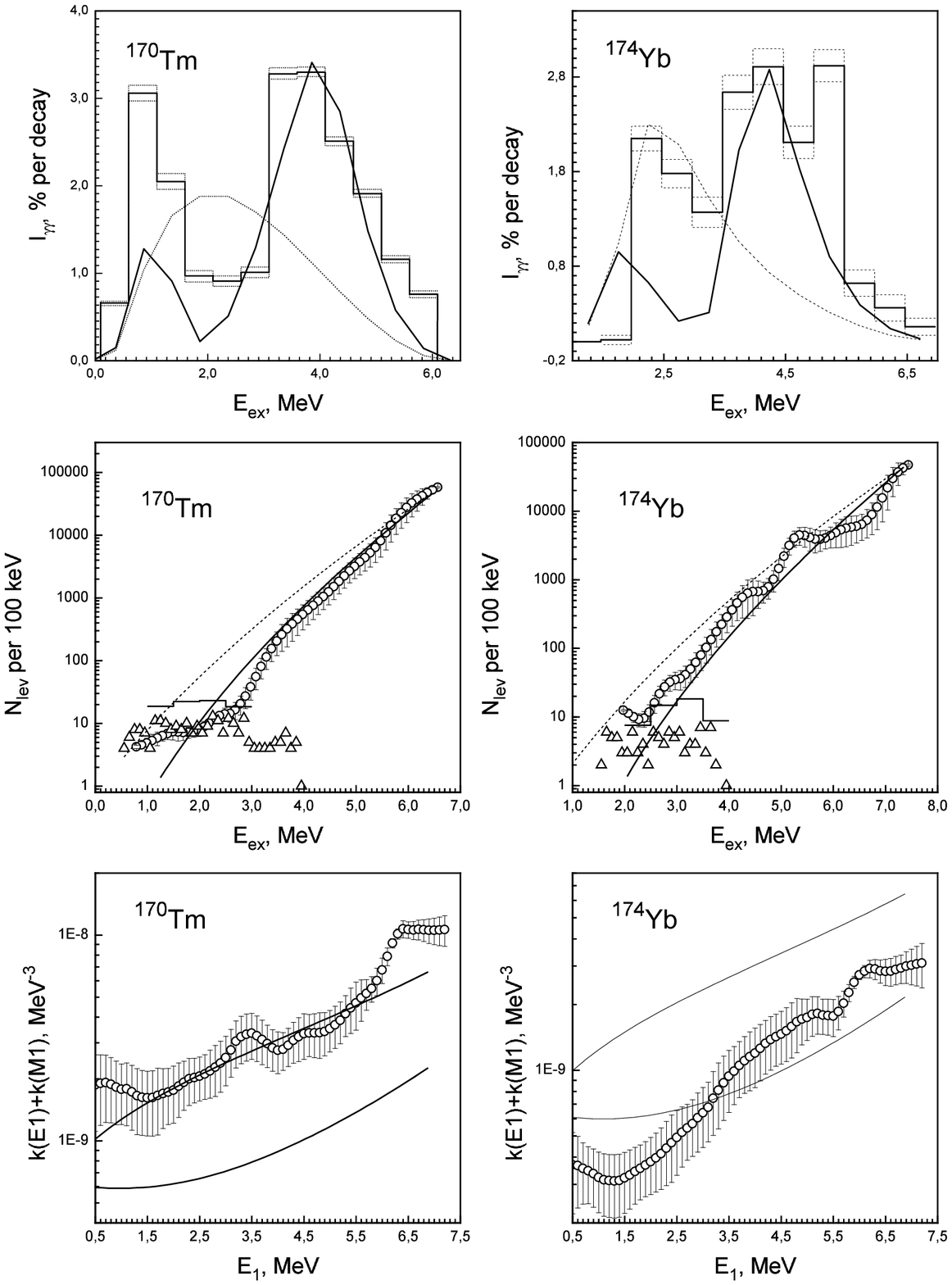}
\end{center}
\hspace{-0.8cm}

{\bf Fig.~10.}~The same as in Fig.~4 for $^{170}Tm$ and $^{174}Yb$.

\end{figure}
\newpage
\begin{figure}
\begin{center}
\leavevmode
\epsfxsize=12.5cm
\epsfbox{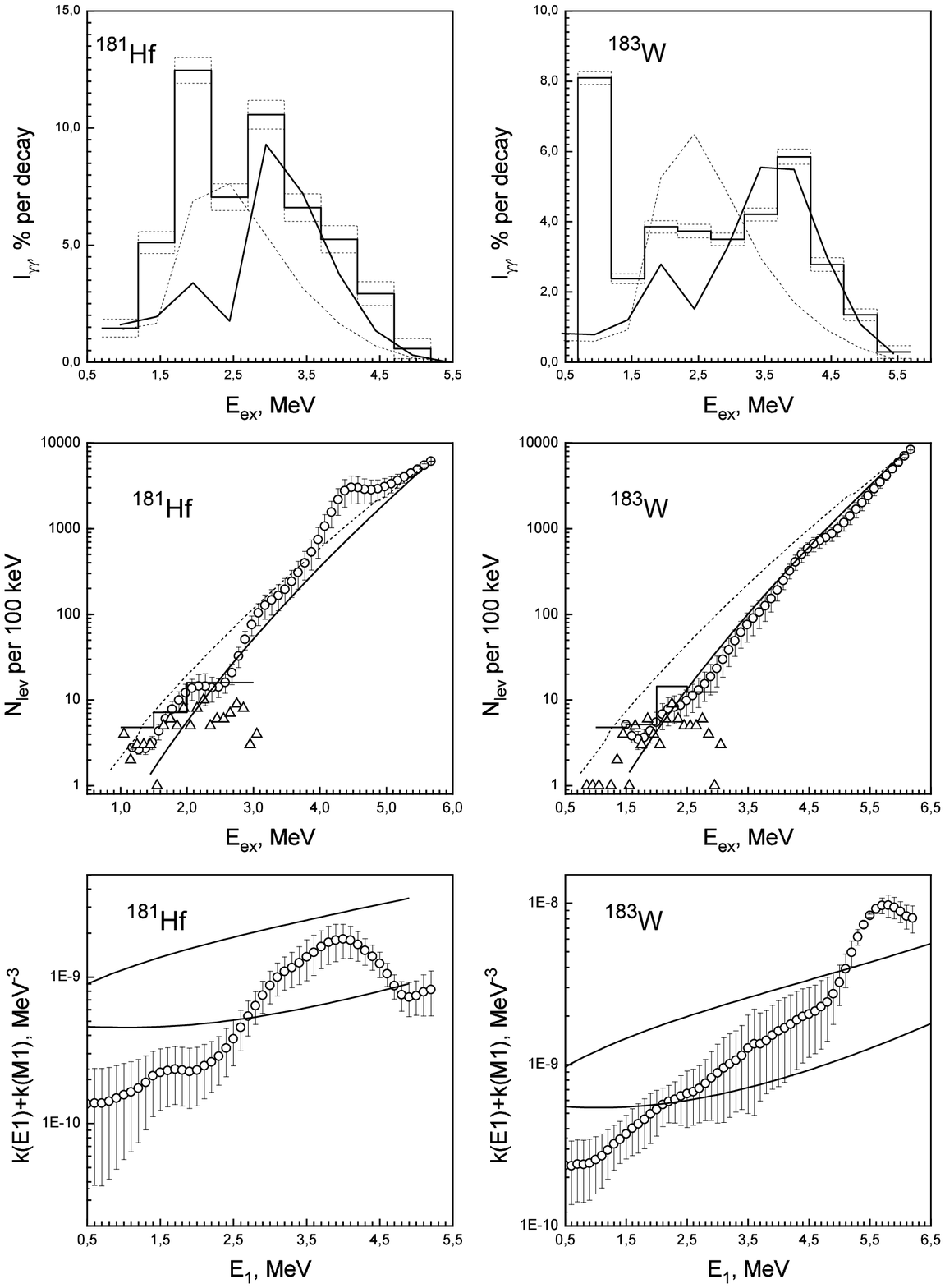}
\end{center}
\hspace{-0.8cm}

{\bf Fig.~11.}~The same as in Fig.~4 for $^{181}Hf$ and $^{183}W$.

\end{figure}
\newpage
\begin{figure}
\begin{center}
\leavevmode
\epsfxsize=12.5cm
\epsfbox{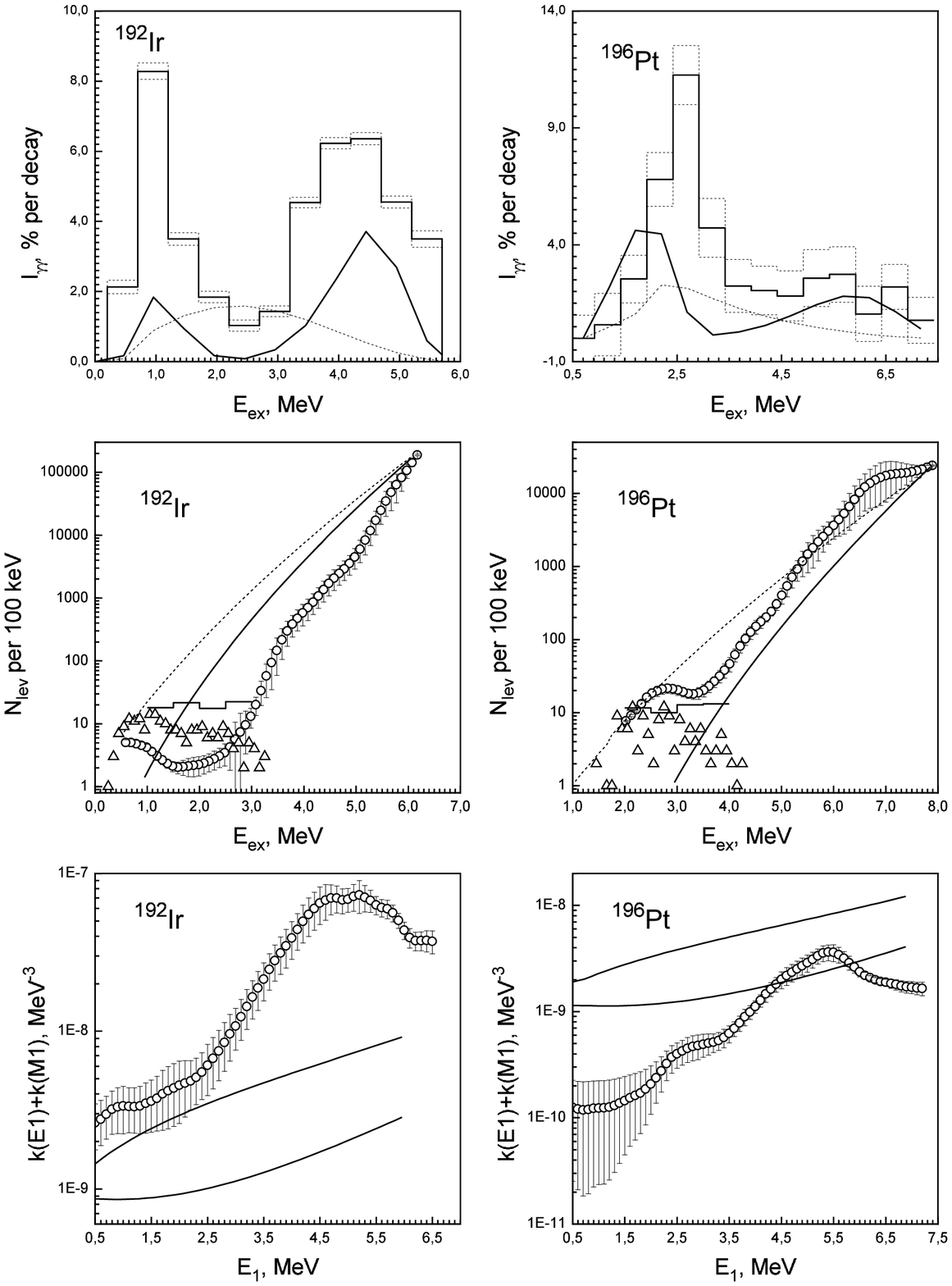}
\end{center}
\hspace{-0.8cm}

{\bf Fig.~12.}~The same as in Fig.~4 for $^{192}Ir$ and $^{196}Pt$.

\end{figure}
\newpage
\begin{figure}
\begin{center}
\leavevmode
\epsfxsize=12.5cm
\epsfbox{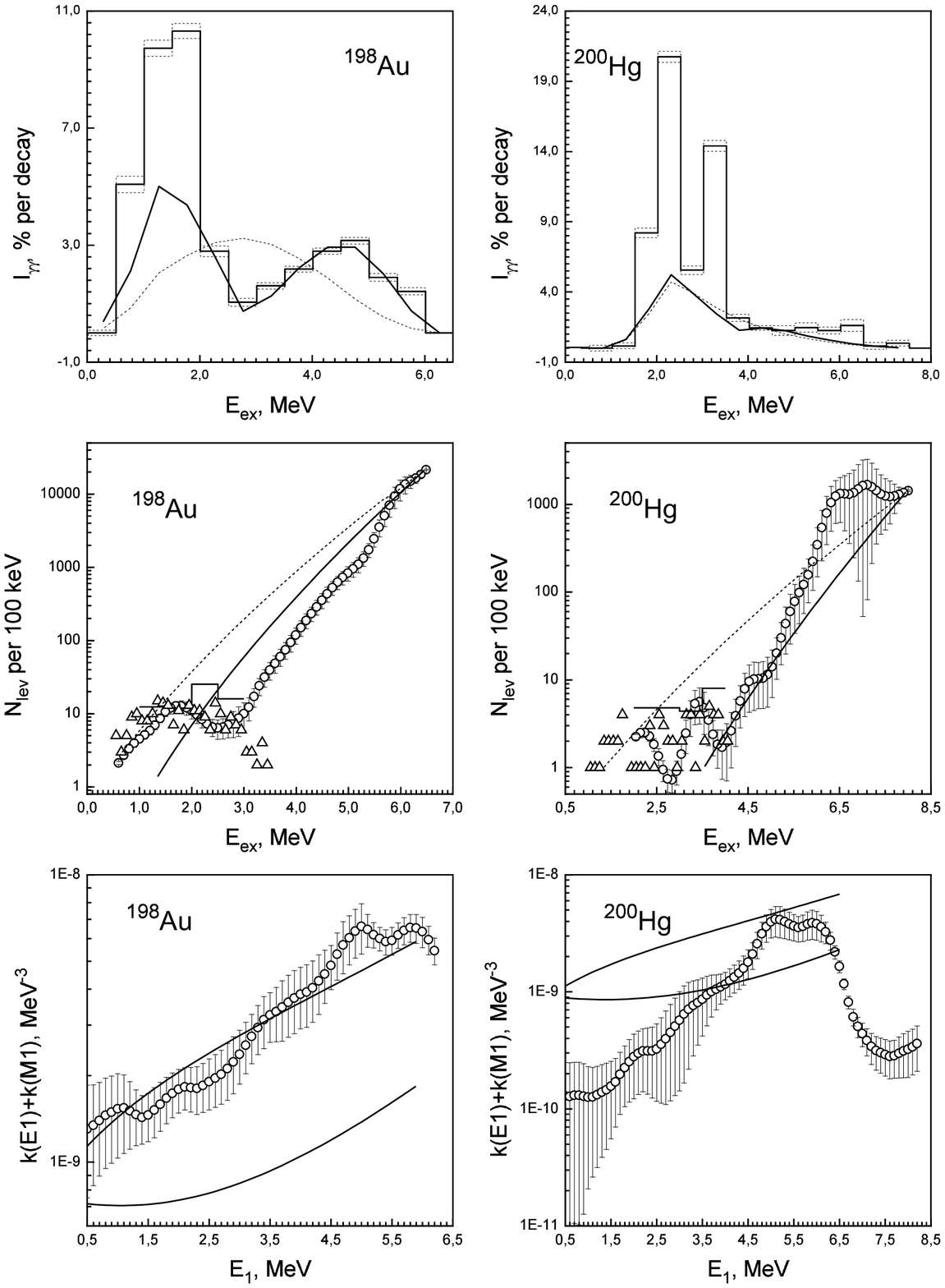}
\end{center}
\hspace{-0.8cm}

{\bf Fig.~13.}~The same as in Fig.~4 for $^{198}Au$ and $^{200}Hg$.
\end{figure}
\end{document}